# Silver Nanowires on Carbon Nanotube Aerogel Sheets for Flexible, Transparent Electrodes

*Patricia M. Martinez* [1, 2], *Arthur Ishteev* [3, 4], *Josef Velten* [1], *Azin Fahimi* [†, 5], *Izabela Jurewicz* [5], *Alan B. Dalton* [6], *Ray H. Baughman,* [1, 2] *and Anvar A. Zakhidov* [*, 1, 3, 4]

1 University of Texas at Dallas, NanoTech Institute, Richardson, TX, 75080, United States.

2 University of Texas at Dallas, Department of Chemistry, Richardson, TX 75080, United states.

3 National University of Science and Technology, MISiS, Leninskiy prospect, Moscow, 119049, Russia.

4 ITMO University, 49 Kronverksky Pr. St. Petersburg, 197101, Russia.

5 University of Surrey, Guildford, Surrey GU2 7XH, United Kingdom.

6 University of Sussex, Falmer, Brighton BN1 9RH, United Kingdom.

ABSTRACT: Flexible, free-standing transparent conducting electrodes (TCEs) with simultaneously tunable transmittances up to 98% and sheet resistances down to 11 Ω/sq were prepared by a facile spray-coating method of silver nanowires (AgNWs) onto dry-spun multiwall carbon nanotube (MWNT) aerogels. Counterintuitively, the transmittance of the hybrid electrodes can be increased as the mass density of AgNWs within the MWNT aerogels increase, however, the final achievable transmittance depends on the initial transparency of the MWNT aerogels. At the same time, a strong decrease in sheet resistance is obtained when AgNWs form a percolated network along the MWNT aerogel. Additionally, anisotropic reduction in sheet resistance and polarized transmittance of AgNWs/MWNTs aerogel is achieved with this method. The final AgNWs/MWNTs hybrid TCEs transmittance and sheet resistance can be fine-tuned by spray-coating mechanisms or by the choice of initial MWNT aerogel density. Thus, a wide range of AgNWs/MWNTs hybrid TCEs with optimized optoelectronic properties can be achieved depending of the requirements needed. Finally, the free-standing AgNWs/MWNTs hybrid TCEs can be laminated onto a wide range of substrates without the need of a bonding aid.

1. Introduction

The need to find materials for replacing Indium Tin Oxide (ITO) in flexible Transparent Conducting Electrodes (TCEs) has stimulated the search for low-cost, lightweight alternatives that have highly transparency and electrical conductivity, high mechanical strength, and compatibility with the numerous substrates used for optoelectronic devices. Though ITO combines high optical transmittance (>90% at 550 nm) with low sheet resistance (10 Ω/sq)[1], applications of ITO-based flexible electrodes are limited by the scarcity of indium and the inherent brittleness of metal oxides[2,3]. Novel technologies, such as for ITO nanowires[4] and ITO mesh structures prepared via photolithography[5], have been developed to prevent ITO from cracking. However, the fabrication of such flexible ITO TCEs requires complex processes that are expensive[1,4]. Efforts have been made to develop flexible TCEs based on multi-walled carbon nanotubes (MWNTs)[6–8], single-walled carbon nanotubes[6,9–16], graphene[17–20], and a combination of the above[3,21–23] as alternatives to ITO[6]. Of particular interest are free-standing MWNT aerogels sheets that can be drawn in the dry state from the side of spinnable MWNT forests[24]. These highly oriented aerogels sheets are porous, transparent, flexible, and conducting. When contacted with organic solvents, such as isopropyl alcohol (IPA) or methanol, they densify during solvent evaporation, reducing sheet thickness from ~20 μm to ~50 nm. MWNT aerogels can also be laminated to a wide range of substrates by contacting the desired section with an organic solvent and allowing surface tension effects during evaporation to densify the aerogel, increasing the contact area between MWNTs and the substrate[24]. However, these self-supported aerogels are highly electronically and optically anisotropic, which can be an undesired property for optoelectronic devices.

Though the use of carbon-based nanomaterials improves the flexibility of TCEs, their low conductivity and transmittance limits optoelectronic performance[25]. Their comparatively high sheet resistances and low transmittances make them unsuitable choices when the need is an optical transmittance above 90% and a sheet resistance below 100 Ω/sq[26,27]. Thus, substantial optical and electrical improvements are necessary for carbon-based nanomaterials to become viable TCEs alternatives.

By integrating silver nanowires (AgNWs) with carbon-based nanomaterials, hybrid TCEs having low sheet resistances and high transmittances, similar to ITO, can be realized[28,29]. The optical and electrical properties of these hybrid TCEs improve when AgNWs percolation networks are formed, increasing the number of conduction pathways connecting AgNWs either to themselves or to the carbon nanomaterial[27,30,31]. These hybrid TCEs provide direct proportionality between sheet resistance and transmittance, as the amount of AgNWs deposited defines the maximum sheet resistance that can be obtained without losing transparency of the substrate [27]. Previous studies have shown that AgNWs can be easily combined with carbon nanomaterials by spray-coating, spin coating, drop-casting, or roll-coating to form hybrid TCEs[3,17,37–45,18,19,21,32–36]. From these techniques, spray-coating is the most suitable and scalable technique since it can be use over large volumes with high reproducibility, while insuring homogeneous AgNW distributions[31,46].

Typically, AgNWs-carbon hybrid TCEs deploy polyethylene terephthalate (PET)[47–50], polyethylene naphthalate (PEN)[51], or polydimethylsiloxane (PDMS)[52] as the flexible substrate and receive a post-treatment to further optimize the transmittance and sheet resistance . Post-treatments ranging from thermal annealing[53,54], mechanical pressing[55], photophysical welding[56,57], low energy plasma[58], or electrochemical Ostwald ripening processes[59] are quite successful, but require fabricating the hybrid electrodes separated from the optoelectronic devices to prevent damage of the layers. Free-standing TCEs that can be laminated onto any device layer have been developed for single-component-based materials, such as metallic AgNWs networks[60], graphene films[61], single-walled carbon nanotubes[62] or MWNT aerogels[24]. In contrast, most of the reported hybrid carbon/metal TCEs are substrate-supported. Hence, it would be valuable to fabricate free-standing AgNWs-carbon based hybrids that meet industrial requirements for TCEs and could be laminated onto any substrate surface.

In this work, free-standing carbon/metal hybrid TCEs are produced in which the optical, electrical and mechanical properties of dry-spun MWNT aerogels are combined with the attributes of AgNWs. To avoid the trade-off between transmittance and sheet resistance, free-standing MWNT aerogels are spray-coated with a solution of AgNWs in IPA. This results in simultaneous increase in transmittance and decrease in sheet resistance as the density of AgNWs increases within the MWNT aerogel. Taking advantage of MWNT aerogel densification by evaporating organic solvents, apertures greater than the wavelength of light form in the aerogel, which increases overall transmittance. Also, relying on the high thermal conductivity of MWNT aerogels[63], the contact resistance of the AgNWs is reduced when a fast 1-2 minute thermal annealing process is performed. As a result, free-standing AgNWs/MWNTs hybrids with sheet resistances ~11 Ω/sq and transmittances ~98% are produced that can be laminated onto targeted substrates. Thus, the described technology produces AgNWs/MWNTs hybrid TCEs that exceed industrial requirements for TCEs in optoelectronic devices.

## 2. Experimental Section

**2.1. Materials**. Two types of AgNWs suspensions, AgNW-25 and AgNWs-60 with concentrations of 5mg/mL in isopropyl alcohol (IPA) were purchased from Seashell Technologies. The mean lengths were 30 μm and 15 μm, with mean diameter of 25 nm and 60 nm, respectively. The AgNWs solution was diluted to a concentration of 0.25 mg/ml with IPA prior to spray coating. AgNWs with diameters between 50-150 nm and lengths greater than 20 μm were synthesized by a polyol method, described by Vinogradov et al [64].

**2.2. Silver evaporation.** A 10 nm thick layer of silver was evaporated onto MWNT aerogels with a CHA-50 electron beam evaporator at a rate of 1.2-1.7 A/sec. Prior to Ag deposition, the MWNT aerogels were vapor-densified with IPA.

**2.3. Synthesis of MWNT aerogels.** Free standing MWNT aerogels were drawn from the sidewall of a spinnable MWNT forest synthesized by chemical vapor deposition (CVD)[24]. The substrates were prepared by depositing 3 nm of iron catalyst using an electron beam (CHA -50) onto Si wafers bearing 100 nm of silicon oxide. A mixture of acetylene (116.8 sccm) and hydrogen (1354 sccm) in a He (2400 sccm) atmosphere were reacted with the substrates at temperatures between 700-730˚C for 5-10 min.

**2.4. Preparation of the AgNWs/MWNT aerogels.** Dry-spun MWNT aerogels were placed on the top of the sheet supports, which had a 1 x 1-inch rectangular aperture in the middle and copper tape electrodes at the edges. Two configurations were prepared: parallel and perpendicular. In the parallel and perpendicular configurations, the MWNT aerogels were placed with the copper electrodes perpendicular to and parallel to the copper electrodes, respectively. Silver paste was painted on top of the contact between the MWNT aerogels and the copper electrodes to ensure sound electrical connection. The sheet resistances, Rs, of the aerogel films were calculated using the formula $Rs = Resistance \left(\frac{MWNT\ aerogel\ lenght}{MWNT\ aerogel\ width}\right)$. The transmittances of the aerogel for 550-nm-light polarized ⊥ and ∥ to the MWNT direction were measured using the polarization mode of a UV-Vis spectrometer (Perkin Elmer Lambda 900 UV-Vis/NIR Spectrophotometer).

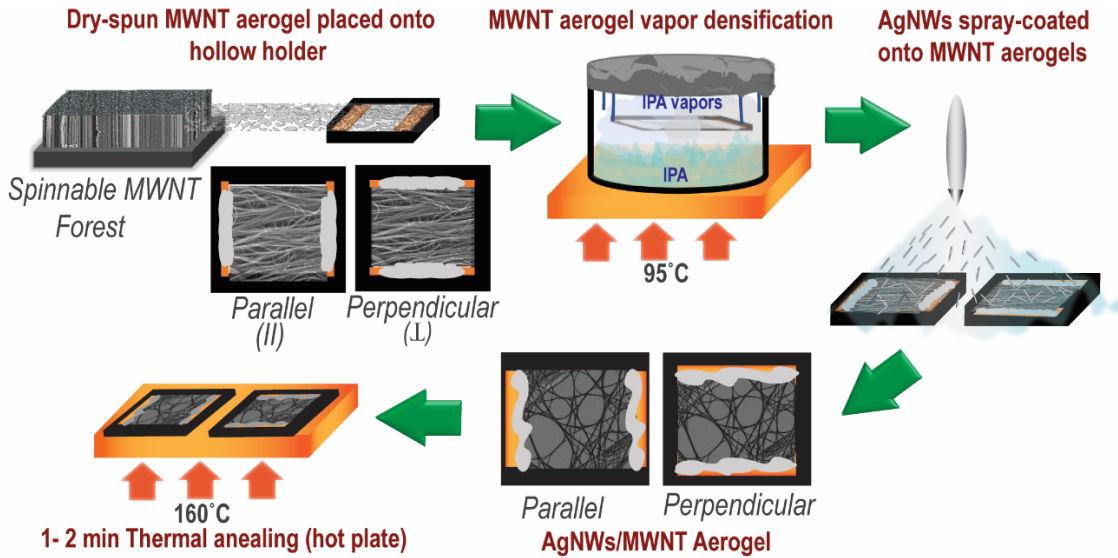

**Figure 1.** Preparation of flexible AgNWs/MWNTs hybrid TCEs

Before spray coating the MWNT aerogels with the AgNW solution, the pristine MWNT aerogels were vapor densified[24] using IPA. To accomplish this, 100 mL of IPA was placed in a 500 mL beaker, covered with aluminum foil, and heated on a hot plate at 95°C for 5 minutes to produce IPA vapors. The aerogels were placed in the beaker (facing down at 2-10 cm distance from the liquid IPA) for 5-30 seconds, and then slowly lifted from the beaker to avoid breakage. A Paasche SI airbrush, which was connected to an Argon or Nitrogen gas source, was used to deposit the AgNWs onto the aerogel at a set pressure of 40 psi. The airbrush was maintained vertically during the entire spray coating process and moved from left to right to ensure uniform coverage of the MWNT aerogel. The distance between the brush nozzle and MWNT aerogels was set at 20-30 cm. Every 20 passes, the sprayed aerogels were placed on a hot plate at 95°C for 5 minutes to complete evaporation of the IPA from the MWNT aerogels. Afterwards, resistance and transmittance measurements were made, as previously described. Thermal annealing at 160°C on a hot plate for 1-2 minutes was performed to decrease the non-effective contacts between AgNWs within the MWNT aerogels and to evaporate any organic solvent residues.

**2.5. MWNT aerogel areal density.** MWNT aerogel areal density was calculated by dividing the weight of one MWNT aerogel sheet by its area. Using measured weights of 4, 6, 8 and 10 MWNT aerogel sheets, linear regression analysis was used to predict the weight of a single aerogel. The weight of the aerogel sheets was an average of three measurements recorded with a microgram accurate scale. To minimize human error, dimensions of the sheets (width and length) were obtained using an optical microscope with attached camera and analyzed using MatLab to obtain an estimation of the area.

**2.6. Physical characterization.** AgNWs/MWNTs aerogels morphologies were characterized using scanning electron microscope, SEM (Zeiss-LEO Model 1530 Variable Pressure Field Effect Scanning Electron Microscope).

**3. Results and discussion**

Figure 2a is a photograph showing the transparency of a AgNWs/MWNTs hybrid TCE that was spray-coated with a 0.25 mg/mL solution of AgNWs and then thermally annealed. This hybrid TCE, which provided T($\perp$)= 98.9% and Rs ($\parallel$)= 11 Ω/sq, is self-supporting and can be transferred to a wide range of substrates (glass, polymer, metal, etc.) without the aid of an adhesive layer. Unless otherwise indicated, the AgNWs/MWNTs hybrid TCEs transmittance is perpendicular transmittance, T($\perp$), taken at 550nm and the sheet resistance is parallel sheet resistance, Rs($\parallel$).

Figure 2b shows the increase in perpendicular transmittance and reduction of parallel sheet resistance, as the amount of deposited AgNWs increases within MWNT aerogels. It highlights the three enhancing mechanisms; Densification Effect, AgNWs Effect and Thermal Annealing Effect, that occur when spray-coating MWNTs aerogels with AgNWs/IPA solution. The first enhancement mechanism, labeled as Densification Effect, is responsible for increasing the AgNWs/MWNTs hybrid TCEs transmittance to values > 90%, and initially reducing the Rs by ~30%. Here AgNWs account for a negligible percentage of the total decrease in Rs since a percolated AgNW system has not been achived[30]. During this first mechanism, fine droplets of AgNWs/IPA locally wet the MWNT aerogel and IPA surface tension forces develop between parallel MWNTs strands, pushing them into thicker bundles with diameters ≤ 5 μm. As the IPA solvent evaporates, strong van der Waals forces develop between individual nanotubes within the bundles allowing them to remain collapsed and create apertures greater than wavelength on light ($\lambda_{\text{visible light}}$), Figure 2c. Simultaneously, the ~30% decrease in aerogel's initial Rs occurs when the solvent densifies, or reduces the aerogel initial thickness from ~20 μm to ~50 nm, creating stronger contacts between individual nanotubes[65]. While the solvent is being evaporated from the MWNT aerogel, AgNWs, carried by the solvent droplets, get deposited on top or between the newly formed MWNT bundles. A significant reduction of aerogel's Rs is only recorded when AgNWs reached the electrical percolation threshold. At this point, AgNWs have form enough connecting paths with themselves, other AgNWs and/or with MWNT bundles through which electrons can travel, Figure 2d.

Once the AgNWs electrical percolation threshold is achieved, the second enhancement mechanism, labeled as AgNWs Effect, takes place and the formation of AgNWs conductive network begins. In this zone, solvent densification increases the transmittance ~0.1% while Rs continues to decrease rapidly as more AgNWs get deposited onto the MWNT aerogel, Figure 2e, reaching a maximum of ~70% decrease of the aerogel's Rs after densification effect.

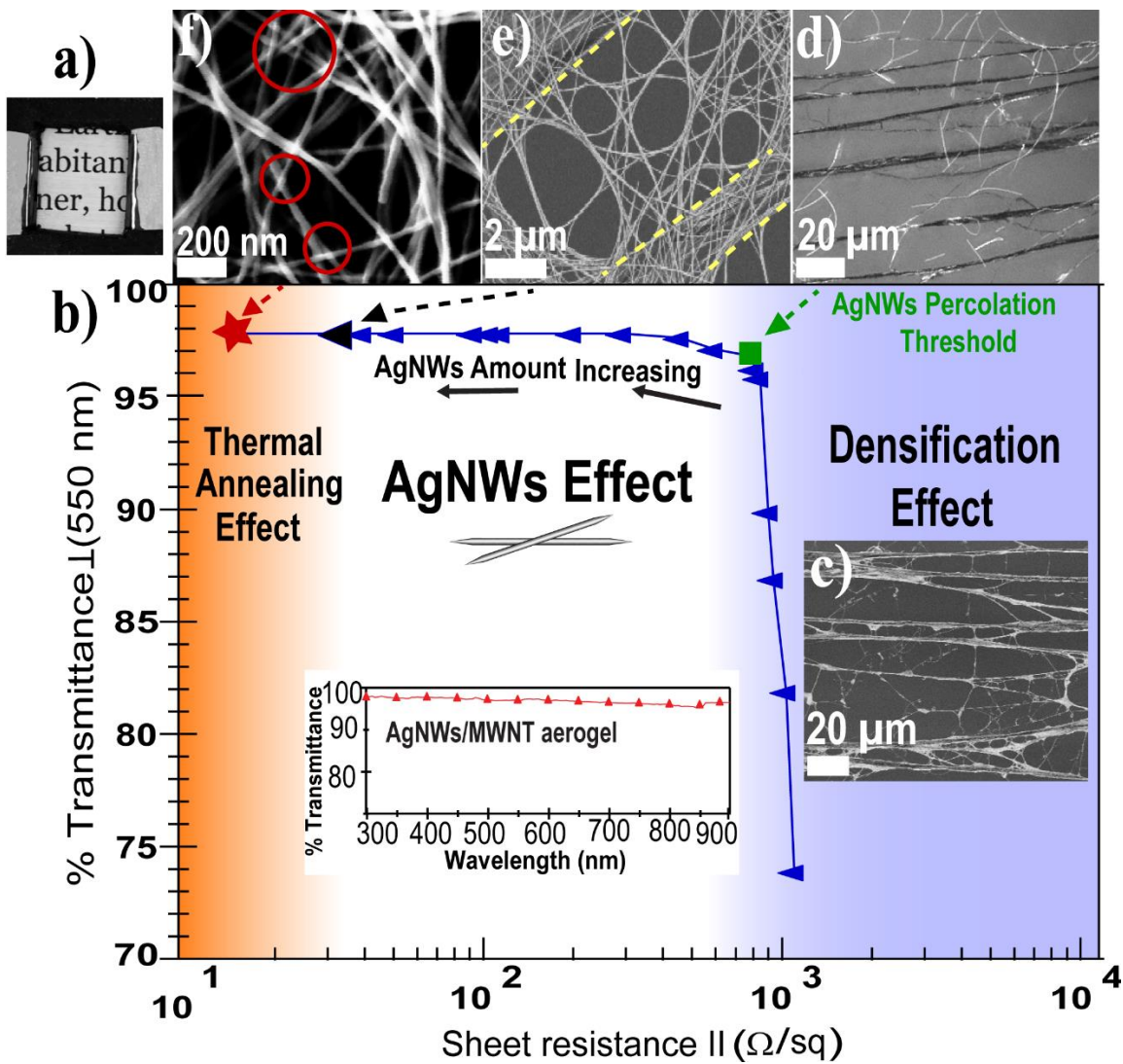

**Figure 2.** a) AgNWs/MWNTs electrode with T(⊥) = 98 % and Rs(∥) = 11.0 Ω/sq. b) The three enhancing mechanisms occurring during spray-coating of AgNWs/IPA to MWNTs aerogels; Densification Effect, AgNWs Effect and Thermal Annealing Effect from which the transmittance is increased while the sheet resistance of the AgNWs/MWNT aerogel is reduced. Inset depicts the optical transmittance of the AgNWs/MWNTs hybrids TCE over the visible spectrum. c) SEM image of MWNT aerogel apertures formed as consequence of the densification effect. d) Optical image representing the AgNWs/MWNT aerogel percolation threshold. e) SEM image depicting the AgNWs/MWNT network taken at the lowest sheet resistance achieved before thermal annealing. Dashed lines highlight the contour of MWNT bundles. f) SEM image of AgNWs welded to one another after thermal annealing.

Once the AgNWs conductive network is achieved within the MWNT aerogel, the Rs is further lowered with the third enhancement mechanism; the Thermal Annealing Effect- a one to two-minute thermal annealing process at 160˚C. This quick annealing takes advantage of the high thermal conductivity of the MWNTs [63] to decrease the non-effective contacts between AgNWs within the MWNT aerogels and evaporates any organic solvent residues from the spray-coating [54], reducing the Rs of the AgNWs/MWNTs hybrids TCEs by 50%. Figure 2f, show the effectiveness of this quick annealing were the AgNWs welded to one another at each contact point without disrupting the MWNT aerogel. Overheating above the ideal temperature or at prolonged times drastically increases the Rs of the AgNWs/MWNTs TCE hybrids due to the disintegration of AgNWs into Ag droplets, (Figure S1).

Dry-spun MWNT aerogels are anisotropic by nature due to the preferential alignment of nanotubes along the pulling direction. Therefore, it is expected that the optical and electrical anisotropic behavior of MWNT aerogels also changes when spray-coated with the AgNWs/IPA solution. A suppression of anisotropic polarized transmittance, TA, along ∥ and ⊥ nanotube direction, is reduced when TA= ∥/⊥ quotient is close to unity. From figure 3a, the AgNWs/MWNTs hybrid TCEs behave as a transparent isotropic material when TA increases from 0.38 to 0.99 as a consequence of MWNT aerogel apertures greater than $\lambda_{visible\ light}$ formed during the Densification Effect.

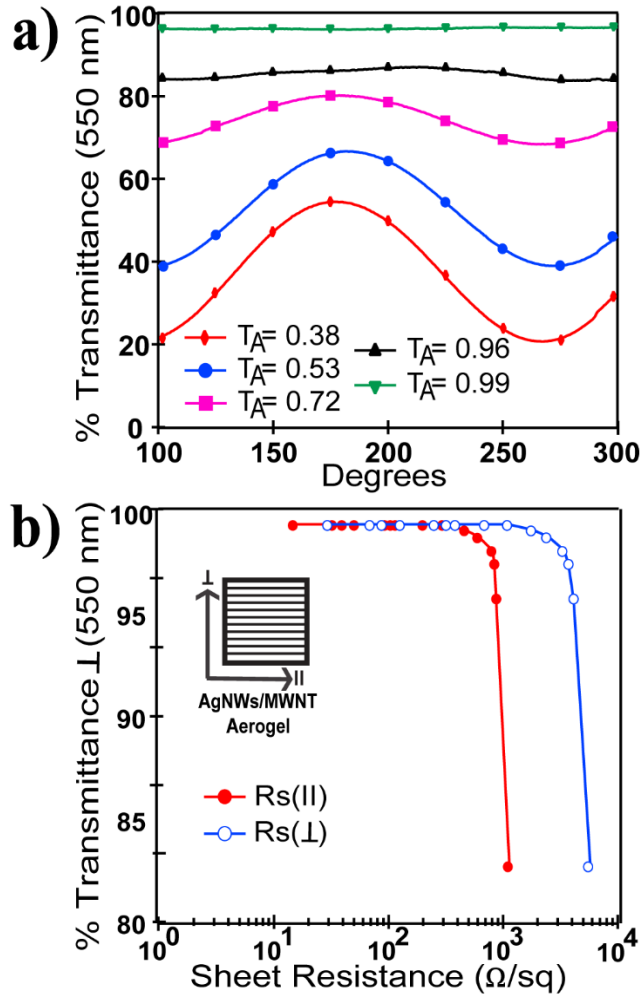

**Figure 3.** a) Suppression of anisotropic transmittance, TA, taken at 550 nm while the light source rotated from 100 to 300 degrees. A maximum transmittance is obtained when $\vec{E}_{field}$ is ⊥ to the MWNTs direction while a minimum is achieved when $\vec{E}_{field}$ is ∥ to the MWNTs

direction. A lost in anisotropic polarization transmittance is achieved when the transmittance anisotropic quotient, TA= ∥/⊥, increased from 0.38 to 0.99, corresponding to the sinusoidal graph profile loss. b) Plot of T(⊥) at 550nm vs Rs in the ∥ and ⊥ MWNTs direction. The Rs anisotropic behavior, $Rs_A$, is reduced when the quotient, $\frac{Rs(\parallel)}{Rs(\perp)}$, increased from 0.20 to 0.40 corresponding to a final Rs(∥)= 11.0 Ω/sq and Rs (⊥)= 29.2 Ω/sq, after annealing treatment.

Rs anisotropic behavior, defined as $Rs_A = \frac{Rs\parallel}{Rs\perp}$, is reduced when the AgNWs electrical percolation threshold is achieved and continues to be reduced until the AgNW conductive network is stablished. A $Rs_A$ quotient close to unity reflects a total loss of anisotropic resistance behavior and represents electrons equally percolating through the AgNWs/MWNT hybrid in the ∥ and ⊥ MWNT direction. Before spray-coating the MWNT aerogels with the AgNWs/IPA solution, the initial aerogel Rs(⊥) is ~70 fold higher than Rs (∥), leading to an $Rs_A$=0.20. After spray-coating 0.180 ± 0.10 µg of AgNWs, the maximum amount needed to achieve the lowest Rs, a ~2.5 fold difference between Rs(⊥) and Rs(∥) is recorded an a $Rs_A$=0.40 is achieved, Figure 3b. Rs(⊥) is the limiting factor in the $Rs_A$ equation. for Rs(⊥) to have a value similar to Rs(∥), AgNWs need to connect adjacent MWNTs bundles and arrange a conductive network in the perpendicular direction, requiring AgNWs with lengths ≥ MWNT aerogel apertures.

### 3.2 AgNWs dimension effects

AgNWs length and diameter affect the onset of the AgNWs electrical percolation threshold, the formation of the AgNW conductive network and thus, the overall Rs of the AgNWs/MWNT hybrids TCEs. Figure 4a, plots the transmittance vs Rs of three types of AgNWs with different diameters and lengths spray-coated onto MWNT aerogels. Also shown in this figure is the effect of spraying exclusively with IPA, outlining the decrease in Rs and increase in transmittance due to densification effect. The highest decrease in Rs is obtained when AgNWs with diameters ~22 nm and lengths ~18 μm are used, Figure 4a, (blue triangles). These AgNWs are flexible enough that, when sprayed, wrap around individual MWNT bundles, creating more opportunities to connect with itself or with other AgNWs, Figure 4c-d, and long enough to form bridges between adjacent MWNT bundles, properties well suited to create the AgNW conductive network. Effectively, this AgNWs reduce the Rs ~ 70% after the densification effect.

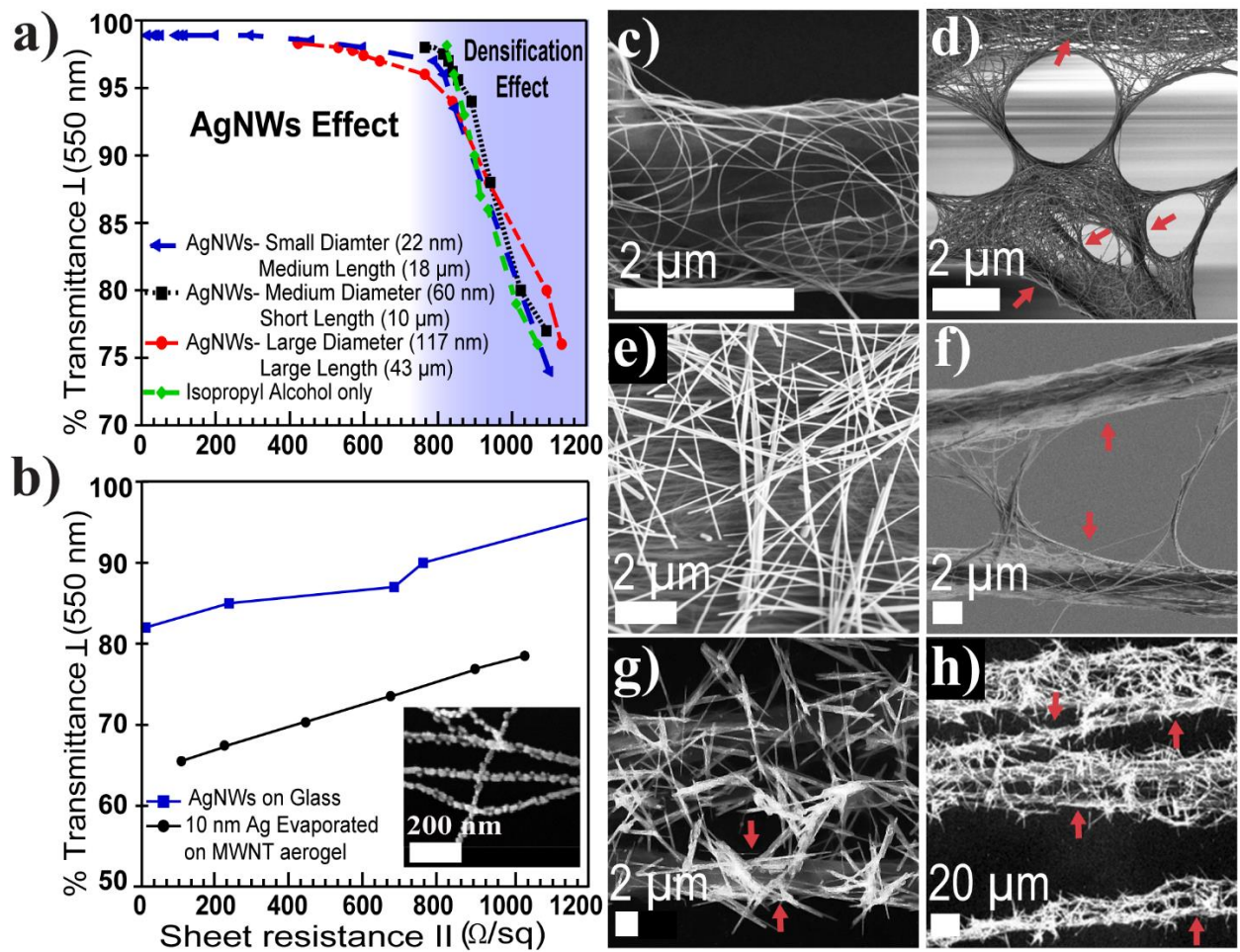

**Figure 4.** a) Evolution of MWNT aerogels increase in transmittance and decrease in Rs when spray-coated with AgNWs with different diameters and lengths. As a reference, the effect of spray MWNT aerogels with IPA is also plotted. Linear decrease in transmittance and Rs is seen when b) 10 nm of Ag are e-beam evaporated onto a MWNT aerogel, black circles, or when AgNWs with small diameters (22 nm) and medium lengths (18 µm), are spray-coated to a glass slide, blue squares. Morphology of MWNT aerogel with 10 nm of Ag evaporated is shown as an inset. SEM of c-d) AgNWs with small diameters (22 nm) and medium lengths (18 µm) wrapping and connecting neighbor MWNT bundles, e-f) AgNWs with medium diameters (60 nm) and short lengths (10 µm) laying on top of a MWNT bundle and forming few bridges between adjacent MWNT bundles. g-h) AgNWs with large diameters (117 nm) and large lengths (43 µm) laying between neighbor bundles without wrapping around them. All the SEM images are taken at the lowest Rs achieved before annealing. Arrows indicate MWNT bundles.

On the contrary, AgNWs with diameters ≥40 nm and short lengths < 10µm, when sprayed to the MWNT aerogels, form non-optimal AgNWs networks and decrease the overall Rs ~10% after densification effect, Figure 4a, black squares. These AgNWs appear to be rigid and lay on top of the MWNT bundles without wrapping around them, forming few bridges between adjacent MWNT when AgNWs agglomeration occur, Figure 4 e-f. If the length of the AgNWs is ≥ 18 µm and the diameter ≥ 40 nm is maintained, a 50% decrease in Rs is seen after densification effect, Figure 4a, red circles. In this case, AgNWs interconnect neighboring MWNT bundles, without wrapping around them, Figure 4 g-f.

Transmittance can still attain high values disregarding the type of AgNWs sprayed onto the MWNT aerogel since it is a process affected solely by MWNT aerogel's solvent densification and independent of AgNWs dimension. The inverse proportionality between transmittance and Rs, is only observed when MWNT aerogels are sprayed-coated with an AgNWs/IPA solution, and not when silver is evaporated onto MWNT aerogels or when AgNWs/IPA are spray-coated to glass slides. Silver evaporated onto a MWNT aerogel is a solvent-free technique and spray-coating AgNWs to glass is unaffected by densification mechanisms thus, transmittance and Rs decrease linearly as the amount of deposited silver increases, Figure 4b.

### 3.3. MWNTs aerogel density effect

MWNT aerogel density plays an important role in constraining the maximum transmittance and minimum Rs, in the $\perp$ and $\parallel$ direction, that the AgNWs/MWNTs hybrids can reach. Ideally, low AgNWs network densities are desired to reduce optical haze of the hybrid TCEs and still obtain low Rs and high transmittances[30]. Though long and thin AgNWs are required to form ideal AgNWs conductive networks necessary to lower the Rs, MWNT aerogel areal density limits the amount of AgNWs per unit area needed to reach the lowest Rs and highest transmittance. Aerogels with high areal densities ~ 5.6 µg/cm$^2$, have initial low transmittances, T= 53 ± 5% and not to high sheet resistances, Rs= 410 ± 25 Ω/sq. These aerogels require a low AgNWs mass density of ~ 20.6 µg/cm$^2$ to obtain a Rs ~10 Ω/sq after annealing, with final T~ 90%. For MWNT aerogels with low areal densities ~2.8 µg/cm$^2$, the initial Rs and transmittances are 825 ± 50 Ω/sq and 70 ± 10%, respectively. With a AgNWs mass density of ~50.0 µg/cm$^2$, these MWNT aerogels can achieve Rs ~11 Ω/sq after annealing; values similar to MWNT aerogels with high areal density, but on the contrary, transmittances as high as 99% can be achieved, Figure 5.

A tradeoff between MWNT aerogel areal density and AgNWs mass density exist that greatly influence the optical and electrical properties of the AgNWs/MWNTs hybrid TCEs. Since low MWNT aerogel areal densities have less CNTs/cm$^2$, when densified with the AgNWs/IPA solution, thinner bundles and bigger apertures are formed allowing the transmittance to achieve values >90%.The lower < 90% transmittance, seen in MWNT aerogels with high areal densities is a consequence of more carbon nanotubes/cm$^2$ available to form additional and/or thicker MWNT bundles when densified, creating apertures lower than the $\lambda_{light}$ that absorb or reflect the incident light more efficiently.

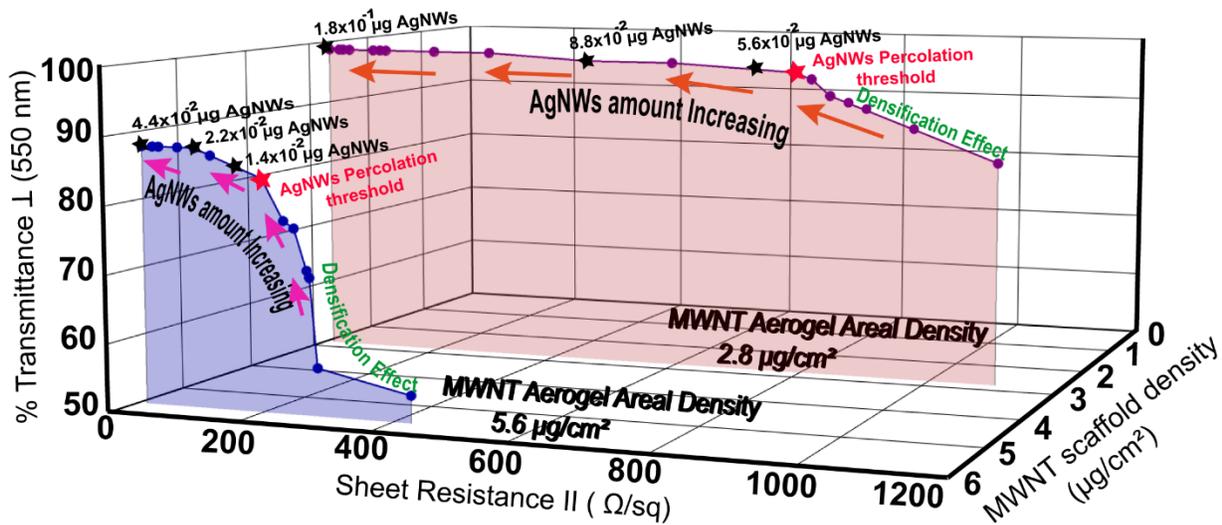

**Figure 5.** Transmittance ⊥ (550 nm) vs sheet resistance ∥ plotted as a function of MWNT aerogel areal density.

### 4. Conclusion

In this study hybrid and self-supported TCEs made of AgNWs spray-coated onto MWNTs aerogels have proven to achieve simultaneous increase in transmittances above 90% and decrease in Rs below 15 Ω/sq by in situ formation of MWNT aerogel apertures and deposition of AgNWs. AgNWs with small diameters (22 nm) and medium lengths (18 μm) were found to create ideal AgNWs conductive networks when spray-coated to MWNT aerogels by wrapping around MWNT bundles, creating effective connections with themselves and with neighbor MWNT bundles. MWNT aerogel Rs and transmittance anisotropic behavior was also suppressed by the formation of apertures with sizes larger than the wavelength of light and by creating new electrical pathways when AgNWs connected neighbor MWNT bundles. The reduced electrical anisotropy is desired for many applications in optoelectronics where charge injection or collection from the transparent electrode is expected to be uniform and isotropic in all direction.

MWNT aerogel with areal densities similar or below 2.8 µg/cm$^2$, were found to be the most suitable aerogels for AgNWs deposition due to their intrinsic high transmittance. However, the high sheet resistance of these low-density aerogels required a greater amount of AgNWs ~50 µg/cm$^2$ to achieve low Rs. Of great interest is the ability to fine tune the amount of AgNWs needed within the MWNT aerogel for achieving the percolation critical point. This suggest further research in achieving an optimal balance of AgNWs and MWNTs possible to maximize the desirable properties of this system as a new form of TCE. What more, these materials can be easily transferred onto various substrates, making them applicable to a wide variety of technologies within the flexible electronics community, allowing for applications not possible with traditional TCEs currently available.

AUTHOR INFORMATION

**Corresponding Author**

* Corresponding author. Tel: 972-883-6218. Email: zakhidov@utdallas.edu (Anvar Zakhidov)

**Present Addresses**

† California Institute of Technology, 1200 East California Boulevard, Pasadena, California 91125, United States.

**Author Contributions**

All authors have given approval to the final version of the manuscript.

**Funding Sources**

Partial financial support from the Ministry of Education and Science of the Russian Federation (Grant № 14.Y26.31.0010 for optical measurements) and grant in the framework of Increase

Competitiveness Program of NUST "MISiS" (No. K2-2015-014 for samples preparation) is acknowledged. Also, a support of Welch Foundation of Texas via grant AT-1617 and Navy STTR Marcorp N181-002-1004 is highly appreciated.

## ACKNOWLEDGEMENTS

Patricia M. Martinez would like to acknowledge CONACYT and the Mexican government for academic opportunities and support.

## ABBREVIATIONS

ITO, Indium Thin Oxide; TCEs, Transparent Conducting Electrodes; MWNTs, Multiwall Carbon Nanotube; AgNWs, silver nanowires; Rs, sheet resistance; T, Transmittance; CVD, Chemical Vapor Deposition.

**Supplementary information**

Thermal annealing at temperatures > 160 ˚C causes the disintegration of AgNWs that connect MWNT bundles and only the AgNWs deposited on top of MWNT bundles suffer a droplet formation, shown in Figure S1. As a consequence, an increase in Rs(∥) is recorded without affecting the T(⊥) at 550 nm.

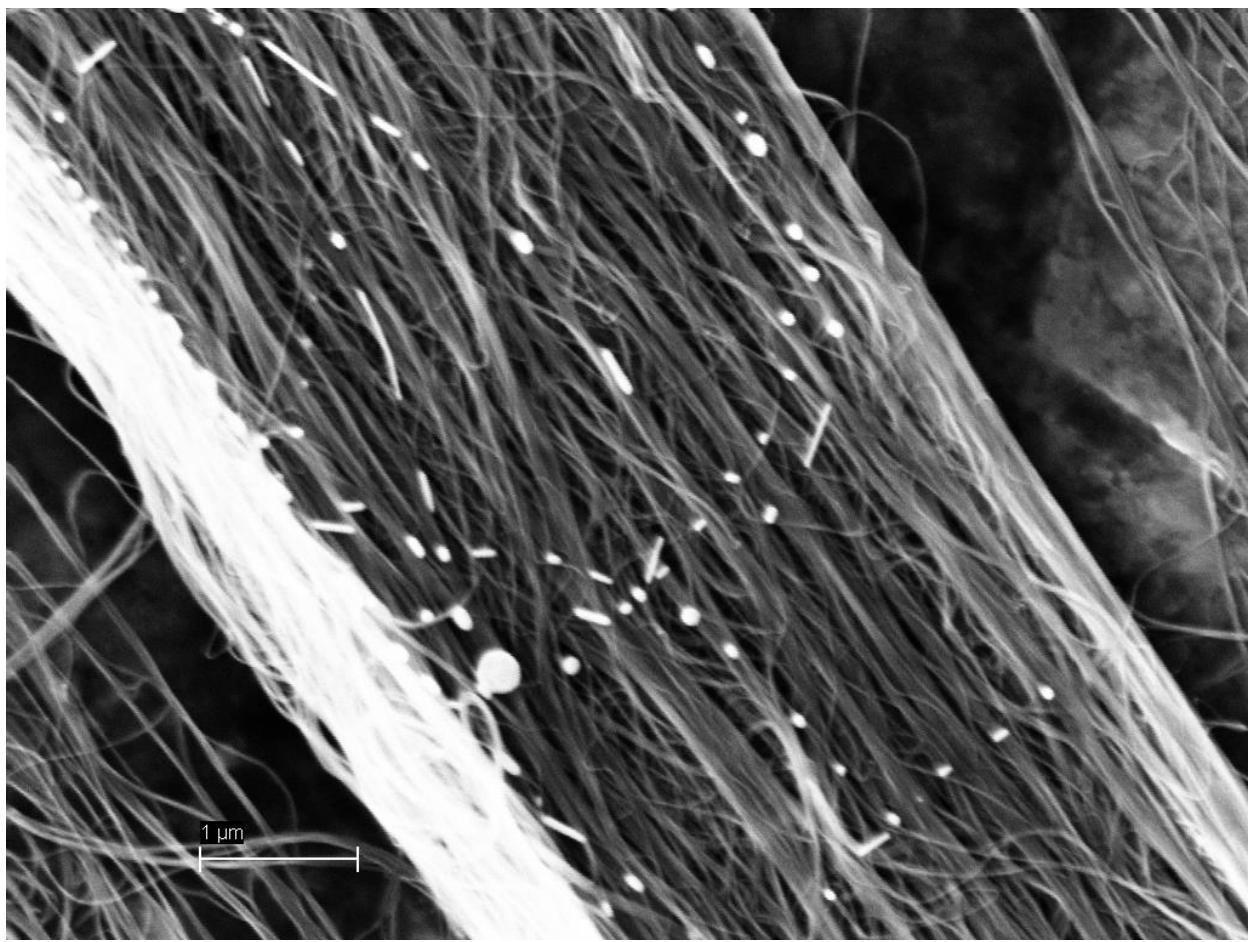

**Figure S3.** AgNWs disintegrate into droplets after thermal annealing treatment at 200˚C.